\documentclass[pageno]{jpaper}

\input{packages}
\newcommand{\ie}{{i.e.}}
\newcommand{\eg}{{e.g.}}

\newcommand{\name}{nanoPU\xspace}

\graphicspath{{figures/}}

\definecolor{LightGray}{gray}{0.9}
\definecolor{Gray}{gray}{0.8}
\definecolor{DarkGray}{gray}{0.5}

\usepackage{listings} % needed for the inclusion of source code

\usepackage{color}

\definecolor{dkgreen}{rgb}{0,0.6,0}
\definecolor{gray}{rgb}{0.5,0.5,0.5}
\definecolor{mauve}{rgb}{0.58,0,0.82}

\usepackage[normalem]{ulem}

\begin{document}

% \title{The \name{}: Making the Network Interface a First-Class Citizen \\ to Minimize RPC Tail Latency \vspace{-0.5em}}
\title{The \name{}: Redesigning the CPU-Network Interface \\to Minimize RPC Tail Latency}

\author{Stephen Ibanez, Alex Mallery, Serhat Arslan, Theo Jepsen, \\Muhammad Shahbaz, Nick McKeown, Changhoon Kim}
\affil{\em Stanford University}

% \affiliation{
%   \institution{\em Stanford University}
% }

\date{}
\maketitle

\thispagestyle{empty}

\begin{abstract}

The \name{} is a new networking-optimized CPU designed to minimize tail latency for RPCs. 
By bypassing the cache and memory hierarchy, the \name{} directly places arriving messages into the CPU register file. 
The wire-to-wire latency through the application is just 65ns, about 13$\times$ faster than the current state-of-the-art.
The \name{} moves key functions from software to hardware: reliable network transport, congestion control, core selection, and thread scheduling. 
It also supports a unique feature to bound the tail latency experienced by high-priority applications.

Our prototype \name{} is based on a modified RISC-V CPU; we evaluate its performance using cycle-accurate simulations of 324 cores on AWS FPGAs, including real applications (MICA and chain replication).
\end{abstract}
\section{Introduction}
% \steve{Chopped out a fair amount of text in the intro and tried to motivate {\em why} tail latency is important.}
% \steve{Maybe we can axe this first paragraph and start with the quote?}
% New distributed applications have demonstrated extraordinary performance running on commodity servers in cloud data centers (\eg, video encoding~\cite{ExCamera}, video compression~\cite{sprocket}, and face recognition~\cite{cirrus}) by dividing computation into fine-grained tasks that execute simultaneously, and by exploiting microservices~\cite{microservices} and serverless computing~\cite{serverless} in software, and RDMA in hardware~\cite{fasst, drtmr, herd}. These applications are typically decomposed into multiple tiers communicating via Remote Procedure Calls (RPCs) and limited by network communication latency. If we can speedup RPCs, then applications can be parallelized further to run even faster.

\epigraph{\vspace{-0.8em}{\small \em ``Sustaining exponential scaling in our computing infrastructure requires much more predictable, low-latency, CPU-efficient communication frameworks starting with RPCs, rather than IP packets.''}}{Amin Vahdat, 2020\vspace{-1em}} 

Cloud service providers (CSPs) are trying to drive down RPC tail latency as more applications are deployed using the microservices architecture~\cite{microservices}.
New distributed applications are demonstrating extraordinary performance running on commodity servers in cloud data centers (\eg, video encoding~\cite{ExCamera}, video compression~\cite{sprocket}, and face recognition~\cite{cirrus}) by dividing computation into fine-grained tasks that execute simultaneously. 
These applications typically fanout Remote Procedure Call (RPC) requests from a root to a large number of leaves, and in multiple tiers. Most often, the service-level performance is limited by the RPC tail latency of individual leaves~\cite{tail-at-scale}. Therefore, if we can reduce (or even bound) RPC tail latency, distributed applications will run faster.

Modern CSPs are attempting to tackle this problem by introducing specialized NIC hardware~\cite{nitro, pensando} with fast RDMA and NIC-resident CPU cores running low-latency microservices.
As a rough rule of thumb, a microservice takes 5--10$\mu$s to invoke and therefore is only worth invoking if we send it for more than $10\mu$s of computation~\cite{shinjuku, shenango, eRPC}.
By comparison, the goal of our work is to enable efficient sub-microsecond RPCs that can be invoked with under $1\mu$s of communication overhead at the server. 
One of the key metrics we use in this paper is the {\em wire-to-wire latency}, defined as the time from when the first bit of an RPC request message arrives at the NIC, until the first bit of the processed RPC response leaves the NIC. 
The best reported median wire-to-wire latency is around 850ns~\cite{eRPC}. 
Our goal is to reduce {\em both} median and tail numbers to below 100ns, making it worthwhile to run ``nanoServices''; short RPCs requiring less than $1\mu$s of work.

Many prior attempts to reduce RPC overhead have included low-latency and lossless switches~\cite{fasst, rdma-shuffle, rdma-networks}, a reduced number of network tiers, and specialized libraries~\cite{eRPC}. 
The current fastest approaches deploy dedicated NIC and switch hardware, but these are hard to program~\cite{hardware-consensus, caribou, netchain, netlock}.

Our work asks the question: {\em Can we design a future CPU core that is easy to program, yet can serve RPC requests with the absolute minimum overhead and tail latency?} 
Our design, which we call the {\em \name{}}, can be seen as a model for future CPU cores optimized for sub-microsecond RPC service, in addition to their regular processing. 
Alternatively, the \name{} can be thought of as a new class of domain-specific nanoService processor, designed to sit on a smartNIC or as a standalone cluster to serve sub-microsecond RPCs. 
For example, it would be practical today to build a single chip 512-core \name{}, similar to Celerity~\cite{davidson2018celerity}, with one hundred 100GE interfaces, servicing over 500 million RPCs per second at a sustained 10Tb/s. 
Such a device could radically improve the performance of large distributed applications. 

Our approach is based on four key observations: First, we need to minimize the time from when an RPC request packet arrives over Ethernet until it starts processing in a running thread.
The \name{} does this by replacing the software thread-scheduler and core-selector ({\em aka} load-balancer) with hardware; by bypassing PCIe, main memory and cache hierarchy completely; by placing RPC data directly into the CPU register file; and by replacing the host networking software stack with a reliable transport layer in hardware, delivering complete RPC messages to the CPU. 
Second, we need to minimize network congestion. 
The \name{} implements NDP~\cite{ndp} in hardware (using a programmable P4 pipeline~\cite{lnic}), reducing congestion and improving incast performance. 
Third, we need to maximize RPC throughput by pipelining header and transport layer processing, thread scheduling and core-selection in hardware. 
The \name{} includes a P4 PISA pipeline~\cite{RMT} in the NIC, processing several packets in parallel, and reassembling RPC messages.
Finally, the performance of large distributed applications is often limited by RPC tail latency; we therefore need to tame and minimize tail latency when processing RPCs.
The \name{} provides what we believe to be the first {\em bounded} tail latency RPC service, guaranteeing that a conforming RPC request will complete service within, for example, $1\mu$s of its arrival at the NIC. 

\subsection{Causes of high RPC tail latency}

\paragraph{a. Memory and cache hierarchy on the critical path.} 
The networking stack of a modern CPU uses memory as a workspace to hold and process packets. 
This inherently leads to interference with applications' memory accesses, introducing resource contention which causes poor RPC tail latency. 
Furthermore, if a packet is transferred over PCIe to DRAM, it is not available to the CPU until several hundred nanoseconds after it arrived~\cite{ramcloud}. 
With direct-cache access technologies (like DDIO~\cite{ddio} and DCA~\cite{dca}), this is reduced, but the packet must still go through many layers of the networking stack, with additional latency for context switching (1--5$\mu$s)~\cite{shinjuku, context-switch-overhead}, including memory copies, TLB flushes, virtual memory management, and cache replacement.

\paragraph{b. Suboptimal scheduling.} When an RPC packet arrives, it must be dispatched to a core for network-stack processing; when complete, the RPC request message is forwarded to a worker core for processing. 
At each step, a software {\em core-selection algorithm}\footnote{Also known as load-balancing or core-steering.} selects the core; and a {\em thread scheduler} decides when processing begins. 
Both algorithms require frequent access to memory by the cores and the NIC, requiring mediation of the memory bus, PCIe, and cache lines. 
Others have shown that these algorithms are on the critical processing path and have attempted to drive down the processing time~\cite{shenango, shinjuku}.
However, the granularity of these software schedulers is inherently limited by the overhead required to perform inter-core synchronization (e.g. sending and receiving interrupts).
Hence, it becomes impractical to make scheduling decisions more than once every $5\mu$s.

% \paragraph{A CPU core processes every packet.} 
% Packet processing cannot be pipelined in software as efficiently as in hardware. Without hardware processing, a CPU core must parse the packet, run a congestion control algorithm, perform message reassembly, then either process the message in place or dispatch it to another core for processing.
% All of these operations must be performed sequentially when executed in software, consuming CPU core time that could be spent processing the RPC request. Recent advances in packet-processing pipelines~\cite{RMT,tofino} suggest these steps can be done faster and more efficiently by a programmable hardware pipeline, with lower latency and higher throughput. 

% In summary, an arriving RPC packet must sequentially traverse a set of layers (namely \emph{header processing}, \emph{core selection}, \emph{thread scheduling}, \emph{transport}, and \emph{memory hierarchy}) before being processed by application code running on a CPU core. 
% Departing RPC responses go through these same layers in reverse. 

While we are not the first to try and reduce the latency of these processing steps~\cite{ramcloud, shinjuku, shenango, eRPC, rpcvalet, nebula}, we believe this paper describes the first complete design to minimize RPC tail latency. 
The \name{} aims to minimize the time spent in every layer, with the judicious use of pipelined, programmable hardware, and direct placement of data into the CPU core---bypassing the memory and cache hierarchy completely.

% \begin{figure}
%   \centering
%   \includegraphics[width=0.70\linewidth]{figures/rpc-pipeline.pdf}
%   \caption{Various stages of an end-host RPC pipeline. \steve{Can probably remove this figure.}}
%   \label{fig:rpc-pipeline}
% \end{figure}

% The approach we take is conceptually simpler than previous work: arriving packets traverse a fixed latency, hardware pipeline until the entire RPC message data (reliably received, and correctly sequenced) is handed to the CPU core for processing, and the CPU thread is invoked within a few nanoseconds.
% Up until this point, the RPC message data has never touched the CPU memory or cache, nor waited to traverse the PCIe bus, nor required any network stack processing in software. 
% A complete RPC message arrives at a core and is made available for clean, continuous reading in the CPU's register file, without any overhead for the core. If the core is ready to begin processing, there is no source of variation in latency, hence creating the opportunity for a {\em bounded} processing time.

\paragraph{Our contributions:}
In summary, we make the following research contributions:
\begin{itemize}
    \item The \name{}: A novel co-design of the NIC and CPU to minimize RPC tail latency. 
    Our design includes: (1) a dedicated memory hierarchy in the NIC, connected directly to the CPU register file, (2) low-latency hardware transport logic, core selection, and thread scheduling, and (3) bounded tail latency by restricting message processing time.
    \item An open-source prototype of \name{}\footnote{The \name{} source code is publicly available at~\cite{nanopu-github}.} extending the RISC\nobreakdash-V Rocket core~\cite{rocket-chip-github} with a 200Gb/s network interface, evaluated with reproducible cycle-accurate simulations on AWS F1 FPGA instances. 
    \item Demonstrated (1) {\em wire-to-wire latency} of just 65ns (13ns without the Ethernet MAC and serial I/0), 13$\times$ faster than the best reported results, (2) 200Gb/s throughput per core, 2.5$\times$ faster than the state of the art, (3) 350 Mpkts/s processing in the NIC (including transport and core selection logic), 50$\times$ faster than the Shinjuku~\cite{shinjuku},  Shenango~\cite{shenango} and eRPC~\cite{eRPC} software solutions, (4) hardware preemptive thread scheduling that enables 99\% tail latency below $2.1\mu$s under high load, (5) the first system to deterministically bound RPC tail latency, and (6) efficient core-selection algorithm in hardware.
    \item Hardware implementation of reliable, low-latency NDP~\cite{ndp} transport layer and congestion control. 
    To our knowledge, it is the first end-to-end evaluation of a hardware transport protocol implemented at an academic institution.
    \item We demonstrate a key-value store (MICA~\cite{mica}) storing 3-way replicated writes in $1.1\mu$s (excluding switch latency), $8\times$ faster than the state of the art~\cite{netchain}.
\end{itemize}

\section{The \name{} Architecture}
\label{sec:design}

\begin{figure*}
  \centering
  \includegraphics[width=0.95\linewidth]{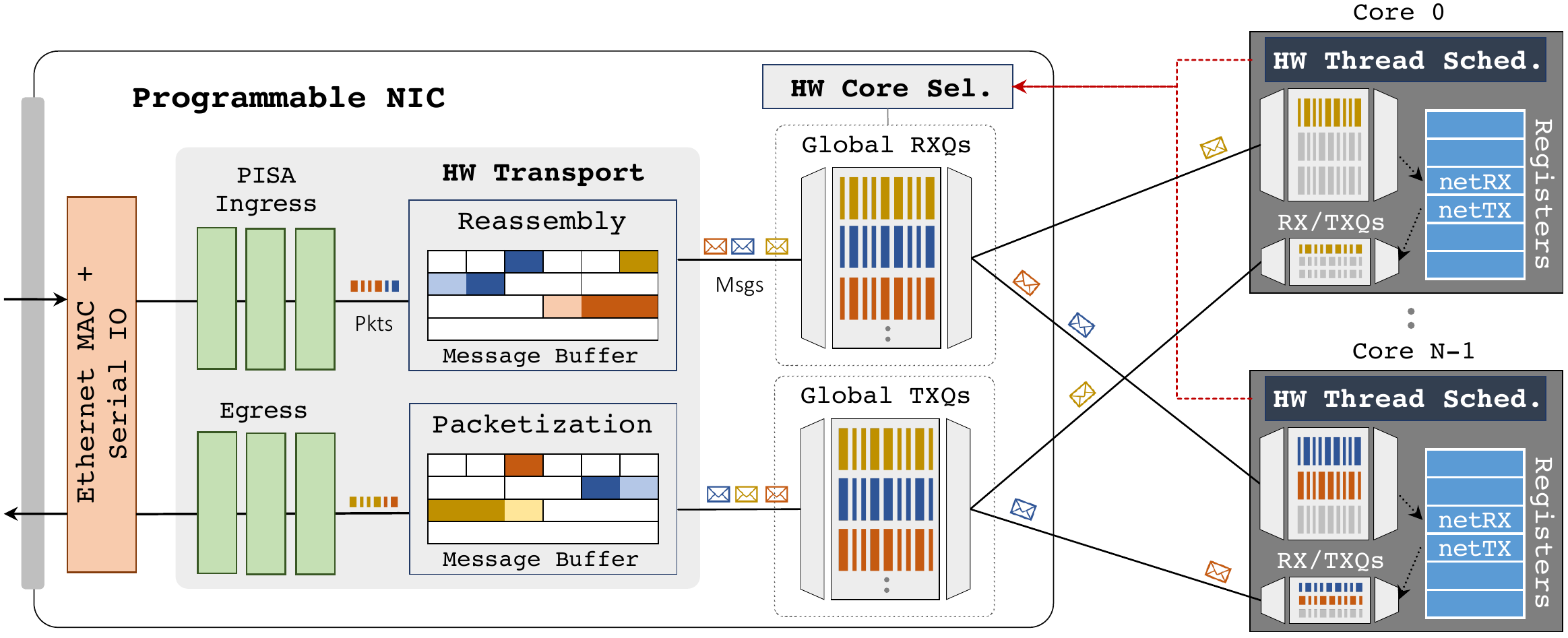}
  \caption{The \name{} design. The NIC includes ingress and egress PISA pipelines as well as a hardware-terminated transport and a core selector with global RX queues; each CPU core is augmented with a hardware thread scheduler and local RX/TX queues connected directly to the register file.}
  \label{fig:nanoPU}
\end{figure*}

Figure~\ref{fig:nanoPU} is a block diagram of the \name{}. Here, we describe each architectural block in turn. 

\subsection{Hardware Terminated Transport Logic}
\label{sec:design-transport}
The transport logic performs three main pipelined tasks in hardware: (1) it processes packet headers, such as VXLAN, overlay tunnels, encryption and decapsulation, (2) converts between Ethernet frames and application messages (\eg, RPC requests and responses), and (3) performs congestion control to reduce in-network latency.

The \name{} transport logic provides the abstraction of reliable one-way message delivery to applications, as opposed to reliable, bi-directional byte stream provided by TCP.
Reliable one-way message delivery can be efficiently implemented in hardware: it only needs to maintain per-message state, rather than per-connection state, so the state can be freed once the message has been successfully delivered to the destination (either a remote host or a local core). Per-message state requirements are small; beyond storing the actual message, we keep a per-message bitmap of received packets, and a few bytes for congestion control. The number of simultaneous RPC communications scales linearly with the number of outstanding messages, rather than the number of hosts in a datacenter, allowing large scale, highly distributed applications involving thousands of \name{}s.
We are not the first to suggest a reliable message abstraction at the transport layer~\cite{homa, ndp}, but we believe we are the first to place it in programmable hardware, allowing multiple transport algorithms to be compared.

A hardware transport layer can be heavily pipelined, allowing it to process several packets at the same time, and freeing the CPU to focus on what it should be running---application code. 
As a secondary benefit, by implementing the transport logic in a fixed latency hardware pipeline, the tail latency of processing each packet is significantly lower than the same algorithm running in software.
Furthermore, a hardware transport layer responds faster than software, leading to a tighter congestion control loop between end-points, and hence more efficient use of the network.

Implementing programmable transport logic in hardware requires support for the following functions in the NIC:
\begin{itemize}[topsep=0.4\baselineskip, leftmargin=20pt]
    \item {\bf Packetization/retransmission buffer} to break a message into packets, and to store outgoing packets until they are acknowledged by a receiver.
    \item {\bf Reassembly buffer} to handle out-of-order packets.
    \item {\bf Timers} and timer-based event processing logic for state transitions, such as retransmissions and background maintenance tasks.
    \item {\bf Schedulers} to decide the order of the outgoing packets. 
    \item {\bf State machines} to maintain per-message state, including the current rate or congestion-window size, sequence and acknowledgement numbers, and message status; and to maintain counters.
    \item {\bf Packet generators} to support transport protocols that generate control packets in response to data-plane events, such as the arrival of a data packet or the detection of a packet drop.
\end{itemize}

The transport logic can be implemented with the help of a P4 programmable, event-driven PISA pipeline~\cite{event-driven-pisa}.
Programmability allows us to compare different transport layers, and allows network owners to create new solutions, possibly dynamically deploying workload-specific transport layers. 
A programmable pipeline means new line-rate packet-processing is easy to add without slowing down application processing in the CPU core, for example network telemetry~\cite{INT} or new protocol headers. 

%%\subsection{Dedicated Memory Hierarchy For the Network}
\subsection{Purpose-built Contention-free IO for the Network}
\label{sec:design-memory}
Recent work has shown that long tail latencies can be caused by bandwidth contention for the main memory; arriving and departing network packet data fights for memory bandwidth with memory accesses by applications~\cite{nebula, resQ}. 
Applications usually process network data sequentially; therefore, random access memory is not the right type of resource for networking in the first place. 
Instead, the \name{} maintains a dedicated two levels of FIFO queues for network data, allowing independent, sequential, non-contending reads and writes, as shown in Figure~\ref{fig:nanoPU}.
On the receive path, the two levels of FIFOs consist of local per-core RX queues and global RX queues shared across cores.\footnote{We think of the per-core local queues as the equivalent of the L1 cache, but for network messages; both are built into the CPU pipeline and sit right next to the register file.}
On the transmit path, there is a corresponding set of local and global TX queues to store individual message words written by applications.

When an application thread running on a core wishes to perform network IO, it binds to a layer-4 port number. 
The \name{} then allocates local and global RX/TX queues for the port.
Threads running on the same core must bind to different port numbers, but threads running on different cores are allowed to bind to the same port number, allowing multiple cores to process messages from the same global RX queue.

The per-core FIFOs connect directly to two general purpose registers (GPRs) in the CPU register file; the head ({\tt netRX}) and tail ({\tt netTX}) of the network RX and TX queues, respectively.
To receive a message, an application simply reads from GPR {\tt netRX}, pulling data from the head of the network RX queue.
Similarly, to send a message, an application writes to the GPR {\tt netTX} corresponding to the tail of the network TX queue.
The hardware ensures that the correct queues are read from and written to for the current thread, preventing data leakage between separate threads.

While the general architecture could be partitioned and packaged in a number of different ways, we assume here that a \name chip contains the NIC and the cores, and that all buffer and FIFO memory is integrated onto the same chip.
Hence, arriving data traverses a dedicated point-to-point link to each core and does not need to wait for a shared PCIe bus.

\subsection{Hardware Core Selection}
\label{sec:design-load-balancing}
Arriving messages must be dispatched to a core for processing by an application thread. 
If the thread is pinned to a single core, the choice is clear. 
But more often, applications run threads on many cores, and we want to dispatch an arriving message to an idle core. 
Ideally, the NIC would maintain a single work-conserving global RX queue from which an idle core can pull its next message to process, leading to the lowest expected waiting time. 
But this design is impractical, requiring all cores to read from a single global RX queue at the same time. 
At the other extreme, where the NIC maintains an RX queue for each core, some messages will become stuck in busy cores' RX queues while other cores are sitting idle. 

Join-Bounded-Shortest-Queue or JBSQ($n$) has been shown to be a good approximation of the ideal single queue system~\cite{r2p2, nebula}, and is practical to implement in hardware. 
JBSQ($n$) uses a combination of a centralized queue, plus short bounded queues of maximum depth $n$ for each core. 
When the per-core queues have available space, the centralized queue will replenish the shortest queue first.
JBSQ(1) is equivalent to the single-queue model.

The \name{} implements the JBSQ($n$) policy in hardware, which maps very naturally onto the two levels of RX queues in Figure~\ref{fig:nanoPU}, with one central RX queue for each layer 4 port number (\ie, application).
By default, we use JBSQ(2), although the design can be configured to use different values of $n$. 
We defer implementation details to Section~\ref{sec:impl-load-balancing}.

\subsection{Hardware Thread Scheduling}
\label{sec:design-thread-scheduling}
The \name{} thread scheduler has several requirements. 
First, it must make decisions frequently and quickly.
The best state-of-the-art operating systems make scheduling decisions every 5$\mu$s~\cite{shinjuku, shenango} making them far too coarse-grained to schedule RPCs with sub-microsecond processing times. 
The \name{} thread scheduler therefore runs in hardware. 
This allows scheduling decisions to be made continuously and in parallel with application processing, and without waiting for a timer interrupt to initiate a context switch to a software scheduler.

Second, the thread scheduler must keep track of which threads are currently eligible for scheduling. 
A thread is marked {\tt active} and therefore eligible for scheduling if the thread has been registered (which means a port number and RX/TX queues have been allocated) and a message is waiting in the thread's local RX queue. 
The thread remains {\tt active} until the thread explicitly indicates that it is {\tt idle} or its local RX queue is empty.

Third, the thread scheduler must choose which thread to run next. 
Each thread is allocated a strict priority; a higher priority {\tt active} thread will preempt a lower priority {\tt active} thread, while threads of the same priority are scheduled to process messages in FIFO order.
As described below, a thread's priority is dynamic: It can be downgraded from priority 0 to priority 1 while running.

Fourth, and unique to the \name{}, the thread scheduler supports guaranteed service time bounds for conforming applications. 
The guarantee, which can only be provided to priority 0 threads, works by limiting their message processing time. 
If a priority 0 thread takes longer than $x$ $\mu$s to process a message, the scheduler will immediately downgrade its priority from 0 to 1, allowing it to be preempted by a different priority 0 thread with pending messages. 
(By default, $x=1\mu$s.) 
If a core is configured to run at most $k$ priority 0 application threads, and these applications are designed such that only one message per application is outstanding at a given moment, then the message processing tail latency for the applications is bounded by: $\max(latency) \le N + kx + (k-1)c$, where $N$ is the NIC latency, and $c$ is the context-switch latency. 
Even if some of the applications on the core misbehave and take longer than $x\mu$s to process a message, this bound still applies to the others.

Finally, the thread scheduler tells the operating system when to change threads. 
It does this by firing an interrupt under the following conditions:
\begin{itemize}
    \item The thread currently running on the core is no longer the highest priority {\tt active} thread. 
    This can happen for a few reasons: (1) a message arrives for a higher priority thread than the one currently running, (2) a high priority thread finishes processing its messages and becomes {\tt idle}, or (3) a priority 0 thread exceeds its maximum allowed processing time and its priority is lowered to 1.
    \item All threads are idle and the current thread exceeds the idle timeout. 
    In this case, the scheduler rotates through all running threads to ensure that they can all make progress.
\end{itemize}

To do its job, the thread scheduler takes three pieces of information as inputs: (1) the state of each thread ({\tt active} or {\tt idle}, (2) the priority of each thread, and (3) the timestamp of the message at the head of each queue.

\begin{figure*}
  \includegraphics[width=0.95\linewidth]{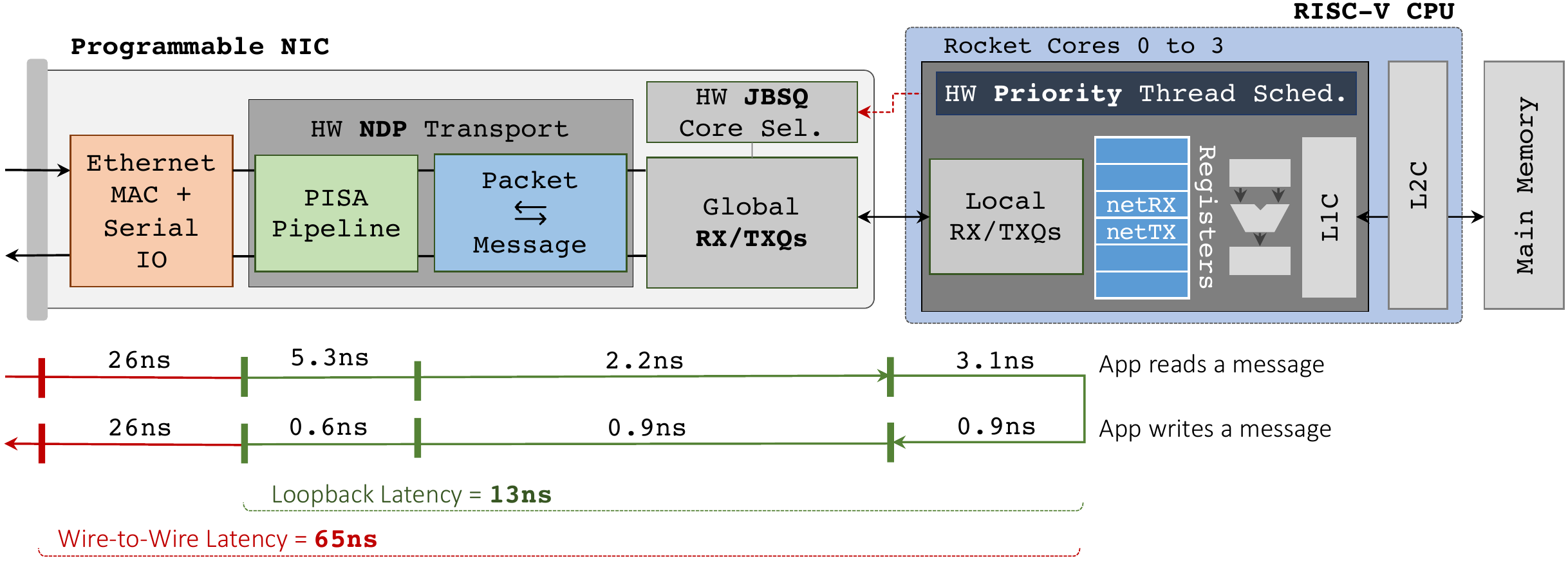}
  \caption{The \name{} prototype and latency breakdown of each stage. Total wire-to-wire latency is {\bf 65ns}.}
  \label{fig:prototype}
\end{figure*}

\section{Our \name{} Implementation}
We designed a prototype quad-core \name{} based on the open-source RISC-V Rocket core~\cite{rocket-chip-github}. 
A block diagram of our prototype is shown in Figure~\ref{fig:prototype}.

Our prototype \name{} extends the open-source RISC-V Rocket-Chip SOC generator~\cite{rocket-chip}, adding 4,300 lines of Chisel~\cite{chisel} code to the code base.
The Rocket core is a simple five-stage, in-order, single-issue processor.
We use the default Rocket core configuration: 16KB L1 instruction and data caches, a 512KB shared L2 cache, and 16GB of external DRAM memory.
Everything shown in Figure~\ref{fig:prototype} except the MAC and Serial IO is included in our prototype and it is available as an open-source, reproducible artifact~\cite{nanopu-github}. 
Our prototype runs a custom-written {\em nanoKernel} based on the RISC-V bare-metal boot code, consisting of about 1,200 lines of C code and RISC-V assembly instructions. 
The nanoKernel is responsible for booting the \name{} and handling the context switch between threads when instructed by the hardware thread scheduler. 
As we will discuss further in Section~\ref{sec:eval}, our prototype \name{} runs on AWS F1 FPGA instances, using the Firesim~\cite{firesim} framework. 

\subsection{Hardware Transport layer}
\label{sec:impl-transport}
\paragraph{NDP: } 
All our evaluations in this paper are based on the NDP transport protocol~\cite{ndp}, which we have implemented on our prototype entirely in hardware. 
We chose NDP because it has promising low-latency performance, is well-suited to handle small RPC messages (the class of messages we are most interested in accelerating), and it can be readily implemented in our programmable pipeline.
See the appendix for a brief summary of NDP functionality.

To evaluate a congested multi-\name{} system, we added NDP's trimming functionality to Firesim's simulated switches. 
We leave the evaluation of our programmable transport layer running other congestion control algorithms to a future paper.

\paragraph{Message buffer: } At 200Gb/s, a new 64B packet can arrive every 2.5ns. 
As a result, in order to run at line-rate, message reassembly and packet retransmission must use very simple operations. 
Three key considerations make it challenging to accomplish this goal: (1) Packets and messages are variable lengths, which makes it difficult to efficiently carve the buffer at high speed; (2) Low-latency transport protocols~\cite{ndp,homa} do per-packet load-balancing, so our design must perform message re-sequencing and reassembly; and (3) Network congestion can cause packet drops, so senders must be able to re-transmit packets out of order. 
The design of our message buffer is simple and not novel, and is described in the appendix.
The key idea is to use fixed size message buffers so that the only operation required to find the position of a packet within a message is to add the appropriate offset to the buffer pointer.

All messages sent and received by applications carry a 64-bit application header, which indicates the message length (in bytes), along with the message's source IP address and source port for received messages; and destination IP address and destination port for transmitted messages.
The transport logic converts between application message headers and the Ethernet, IP, and NDP headers on each packet.

% \begin{figure}
%   \centering
%   \includegraphics[width=0.85\linewidth]{./figures/app-headers}
%   \caption{Format of message headers to/from application threads.}
%   \label{fig:app-headers}
% \end{figure}

\subsection{JBSQ Core Selection in Hardware}
\label{sec:impl-load-balancing}
As explained above, the \name{} implements the JBSQ(2)~\cite{r2p2} core selection algorithm. 
JBSQ(2) is implemented using two tables.
The first maps the layer 4 port number to a per-core bitmap, indicating whether or not each core is running a thread bound to the port number.
The second maps the layer 4 port number to a count of how many messages are outstanding at each core for the given port number.
When a new message arrives, the algorithm checks if any of the cores that are running an application thread bound to the destination port are holding fewer than two of the application's messages. 
If so, it will immediately forward the message to the core with the smallest message count.
If all target cores are holding two or more messages for this port number, the algorithm waits until one of the cores indicates that it has finished processing a message for the destination port. 
It then forwards the next message to that core.

\subsection{Priority Thread Scheduling in Hardware}
The \name{} implements thread scheduling in hardware, as described in Section~\ref{sec:design-thread-scheduling}. 
Our prototype supports running up to four threads on each core; each thread can be configured with a unique priority value.
Priority 0 has a configurable maximum message processing time in order to support bounded service-time applications. 
We added a new {\em thread-scheduling interrupt} to the RISC-V core, along with an accompanying control \& status register (CSR) set by the hardware scheduler that tells the nanoKernel trap handler which thread it should run next. 
When processing very low latency RPCs, we disable all other interrupts to avoid unnecessary traps and context switches.

We define the context-switch latency to be the time from when the scheduler fires the interrupt to when the first instruction of the target thread is executed.
Our prototype has a measured context-switch latency of 160 cycles, or 50ns for a 3.2 GHz CPU.

\subsection{Register File Interface}
The RISC-V Rocket core requires surprisingly few changes to support our model of packets entering and leaving via two reserved GPRs. 
The main change, naturally, involves the register read-write logic. 
Each core has 32 GPRs, each 64-bits wide, and we have reserved two for network communications. 
Applications must be compiled to avoid using the reserved GPRs for temporary storage. 
Fortunately, {\tt gcc} makes it easy to reserve registers via command-line options~\cite{gcc-options}.

The core also requires changes because of pipeline flushes. 
A pipeline flush can occur for a number of reasons (\eg, a branch misprediction).
On a traditional five-stage RISC-V Rocket core, architectural state is not modified until an instruction reaches the write-back stage (stage 5).
However, with the addition of our network register file interface, reading the GPR corresponding to the network RX queue now causes a state modification in the decode stage (stage 2).
This destructive read operation must be undone when there is a pipeline flush. 
In our prototype, the CPU pipeline depth is an upper bound on how many read operations need to be undone; in our case, at most two reads require undoing. 
Fortunately, it is easy to implement a FIFO queue that supports this operation.

\subsection{The \name{} Hardware/Software Interface}
To illustrate the mechanics of how software on the \name{} interacts with the hardware, Listing~\ref{lst:asm} shows a simple loopback-and-increment program in RISC-V assembly.
The program continuously reads 16B messages (two 8B integers) from the network, increments the integers, and sends the messages back to their sender.
The program details are described below.

The \verb|entry| procedure binds the thread to a layer 4 port number at the given priority level by first writing a value to both the \verb|lcurport| and \verb|lcurpriority| CSRs, then writing the value 1 to the \verb|lniccmd| CSR.
The \verb|lniccmd| CSR is a bit vector used by software to send commands to the networking hardware; in this case, it is used to tell the hardware to allocate both local and global RX/TX queues for port 0 with priority 0.
The \verb|lniccmd| CSR can also be used to unbind a port or to update the priority level.

The \verb|wait_msg| procedure waits for a message to arrive in the local RX queue by polling the \verb|lmsgsrdy| CSR until it is set by the hardware.
While it is waiting, the application tells the hardware thread scheduler that it is idle by writing to the \verb|lidle| CSR during the polling loop.
The scheduler uses the idle signal to evict idle threads in order to schedule a new thread that has messages waiting to be processed.

The \verb|loopback_plus1_16B| procedure simply swaps the source and destination addresses by moving the RX application header (the first word of every received message - see Section~\ref{sec:impl-transport}) from the {\tt netRX} register to the {\tt netTX} register, shown on line 19 (Listing~\ref{lst:asm}). 
It then increments every integer in the received message and appends them to the message being transmitted.
After the procedure has finished processing the message, it tells the hardware scheduler it is done by writing to the \verb|lmsgdone| CSR.
The scheduler uses this write signal to reset the message processing timer for the thread. 
It may also evict the thread to ensure that messages arriving for other threads of the same priority are processed in FIFO order.
Finally, the procedure waits for the next message to arrive.

Applications that use variable-length messages can use the message length (in the RX application header) to read the correct number of words from the network RX queue.
If an application reads an empty RX queue, the resulting behavior is undefined---similar to reading an uninitialized variable.

\begin{figure}
  \centering
  \small
  \begin{minipage}[c]{0.9\linewidth}
\begin{lstlisting}[language=riscv]
// Simple loopback & increment application
entry:
  // Register port number & priority with NIC
  csrwi	lcurport, 0
  csrwi	lcurpriority, 0
  csrwi	lniccmd, 1

// Wait for a message to arrive
wait_msg:
  csrr	a5, lmsgsrdy
  bnez	a5, loopback_plus1_16B
idle:
  csrwi	lidle, 1 // app is idle
  csrr	a5, lmsgsrdy
  beqz	a5, idle

// Loopback and increment 16B message
loopback_plus1_16B:
  mv netTX, netRX // copy app hdr from rx to tx
  addi netTX, netRX, 1 // send word one + 1
  addi netTX, netRX, 1 // send word two + 1
  csrwi	lmsgdone, 1 // msg processing complete
  j wait_msg // wait for the next message
\end{lstlisting}
  \end{minipage}
  \vspace{-5pt}
  \captionof{lstlisting}{Loopback with increment. A nanoPU RISC-V assembly program that waits for a 16B message to arrive, increments each word, and returns it to the sender.}
  \label{lst:asm}
\end{figure}
\section{Evaluation}
\label{sec:eval}
We evaluate our \name{} design using micro-benchmarks for latency and throughput, for thread scheduling, load balancing, and congestion control. 
We run real application benchmarks for the MICA key-value store~\cite{mica} and the NetChain chain-replication protocol~\cite{netchain}.

\subsection{Methodology}

\paragraph{Benchmark tools:}
For basic latency and throughput micro-benchmarks, we use the Verilator~\cite{verilator} cycle-accurate software simulator. 
For all other evaluations, we use Firesim~\cite{firesim} to run our design on F1 FPGA instances in AWS~\cite{f1}, allowing us to run large-scale cycle-accurate simulations of applications using hundreds of \name{} cores.
The FPGAs run at 90~MHz, and we simulate a target clock rate of 3.2~GHz---all reported results are in terms of this target clock rate. 
The FPGAs are connected by C++ switch models running on the AWS x86 host CPUs.

\paragraph{Custom load generation in Firesim:}
To evaluate our system's tail latency under load, we added a custom (C++) load generator to Firesim, connected to the \name{} by a simulated network link with 43ns latency. 
In our runs, it generates 20k requests with Poisson inter-arrival times, and measures the end-to-end latency of each RPC call.

\subsection{Microbenchmarks}

\paragraph{a. Wire-to-wire and loopback latency:}
Figure~\ref{fig:prototype} shows the latency breakdown for a single 8B application message (in a 72B packet) measured from the Ethernet wire through a simple loopback application in the core, then back to the wire.\footnote{Our prototype does not include MAC \& Serial IO, so we add real values measured on a 100GE switch (with Forward Error Correction disabled).}
As shown, the loopback latency through the \name{} is only 13ns, but in practice we also need an Ethernet MAC and serial I/O, leading to a wire-to-wire latency of 65ns.
The wire-to-wire latency is about 13$\times$ faster than the current state-of-the-art on a commodity server, eRPC~\cite{eRPC}, which reports a host-stack latency of 850ns.

%\begin{table}[]
%\begin{tabular}{c|c}
%\textbf{Logic}                   & \textbf{Latency (ns)} \\ \hline
%RX MAC \& Serdes                  & 26                                         \\ \hline
%Ingress CC                       & 5.3                                          \\ \hline
%Msg assembly \& delivery to core & 2.2                                          \\ \hline
%App reads msg                    & 3.1                                          \\ \hline
%App writes msg                   & 0.9                                          \\ \hline
%Packetization logic              & 0.9                                          \\ \hline
%Egress processing                & 0.6                                          \\ \hline
%TX MAC \& Serdes                  & 26                                         \\ \hline
%\textbf{Total}                   & \textbf{65}                               
%\end{tabular}
%\caption{Breakdown of the \name{} loopback latency. \xxx{Move these numbers into the figure.}}
%\label{tab:latency}
%\end{table}

\begin{table}[]
\begin{center}
\begin{tabular}{crr}
\textbf{Msg. Length}  & \textbf{RX (Gb/s)} & \textbf{TX (Gb/s)} \\ \toprule
Fixed    & 195                & 200                \\ \hline
Variable & 68                 & 71                
\end{tabular}
\caption{RX/TX throughput of a single-core \name{} for two applications processing 1KB messages: one designed for fixed-length messages and the other for variable-length messages.}
\label{tab:throughput}
\end{center}
\end{table}

\paragraph{b. Single core throughput:}
Table~\ref{tab:throughput} shows the maximum sustainable RX and TX throughput for a single \name{} core, processing 1KB messages for two applications: one designed to process fixed-length messages and another designed to process variable-length messages.
With fixed-length message processing, the send and receive loops can be unrolled, making them three times faster than for variable-length message processing. 
With loop unrolling, almost all instructions perform network reads and writes, whereas without it, 66\% of the instructions are needed to manage the loop (\ie, branch and increment instructions).
eRPC~\cite{eRPC} reports a per-core goodput of up to 75~Gb/s, corresponding to a wire rate of about 78~Gb/s, about 2.5 times slower than the \name{}.

The \name{}'s programmable NIC is designed to process packets at a line-rate of 200~Gb/s.
Thus, for small 8B RPC request messages (transported by 72B Ethernet packets), the NIC supports a maximum throughput of 350 million requests per second (Mrps), or about 50$\times$ higher than existing systems that perform network packet processing and message load balancing in software on a dedicated CPU core~\cite{shinjuku, shenango, eRPC}.

\paragraph{c. Thread scheduling:}
We evaluate the performance of \name{}'s hardware thread scheduler (which has its own interrupt) against a more traditional timer-interrupt driven scheduler.
In both cases, scheduling decisions are made in hardware.\footnote{A software scheduler would either need to make scheduling decisions on a separate core or upon handling the timer interrupt. 
Hence, its performance would only be worse than what we evaluate here.}  
For the timer-interrupt driven thread scheduling policy, we disable the hardware thread scheduler's interrupt and instead configure a timer interrupt to fire every 5$\mu$s, at which point the kernel swaps in the highest-priority active thread. 
We use 5$\mu$s timer interrupts to match the granularity of state-of-the-art low latency operating systems~\cite{shinjuku, shenango}.

We evaluate both schedulers when they are scheduling two threads: one with priority 0 (high) and one with priority 1 (low). 
We tell the load generator to generate requests with an on-core service time of 500ns (\ie, an ideal system will process 2Mrps).

\begin{figure}
  \includegraphics[width=\linewidth]{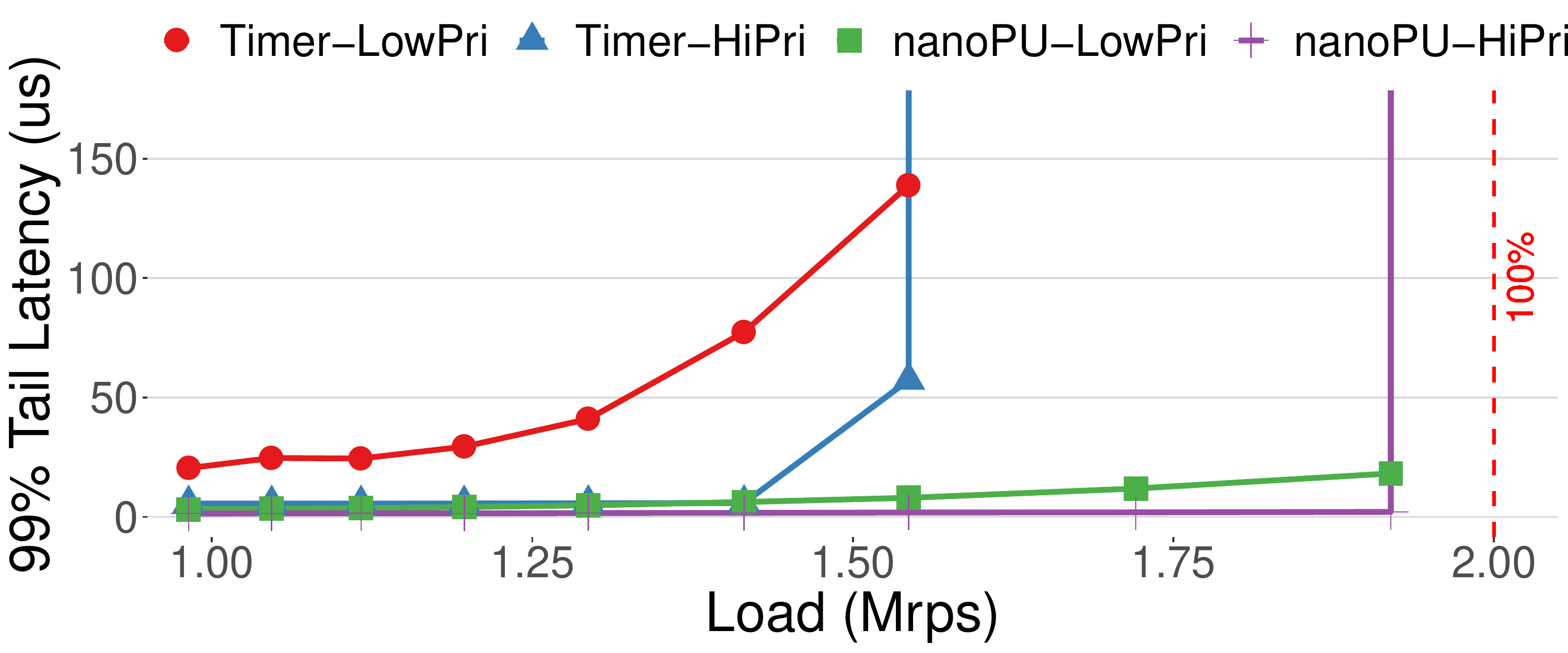}
  \caption{Comparing \name{}'s hardware thread scheduler performance against a more traditional timer-interrupt driven thread scheduler. Graph plots 99\% tail latency vs load for both a high-priority and low-priority thread for each experiment.}
  \label{fig:prio-scheduling-eval}
\end{figure}

Figure~\ref{fig:prio-scheduling-eval} shows the 99\% tail latency vs load for both thread scheduling policies, with a high and low priority thread.
By allowing the hardware to drive the thread scheduling logic as messages arrive, the tail latency of the high and low priority threads are reduced by 4$\times$ and 6.5$\times$ at low load, respectively; and it can sustain at least 96\% load.\footnote{The \name{} does not currently allocate NIC buffer space on a per-application basis. This means that when the RX queue for a low priority application builds up, it can cause high-priority requests to be dropped. This will be improved in the next version of the \name{}.}

\paragraph{d. Bounded message processing time:}
We evaluate the ability of the \name{} to bound the tail latency of well-behaved applications, even when they are sharing a core with misbehaving applications.
To do this, we configure one of the \name{}'s cores to run two threads, one well-behaved thread and one misbehaving thread.
All requests have an on-core service time of 500ns, except when a thread misbehaves (once every 100 requests), in which case the request processing time is $5\mu$s.
Both threads are configured to run at priority 0.

\begin{figure}
  \includegraphics[width=\linewidth]{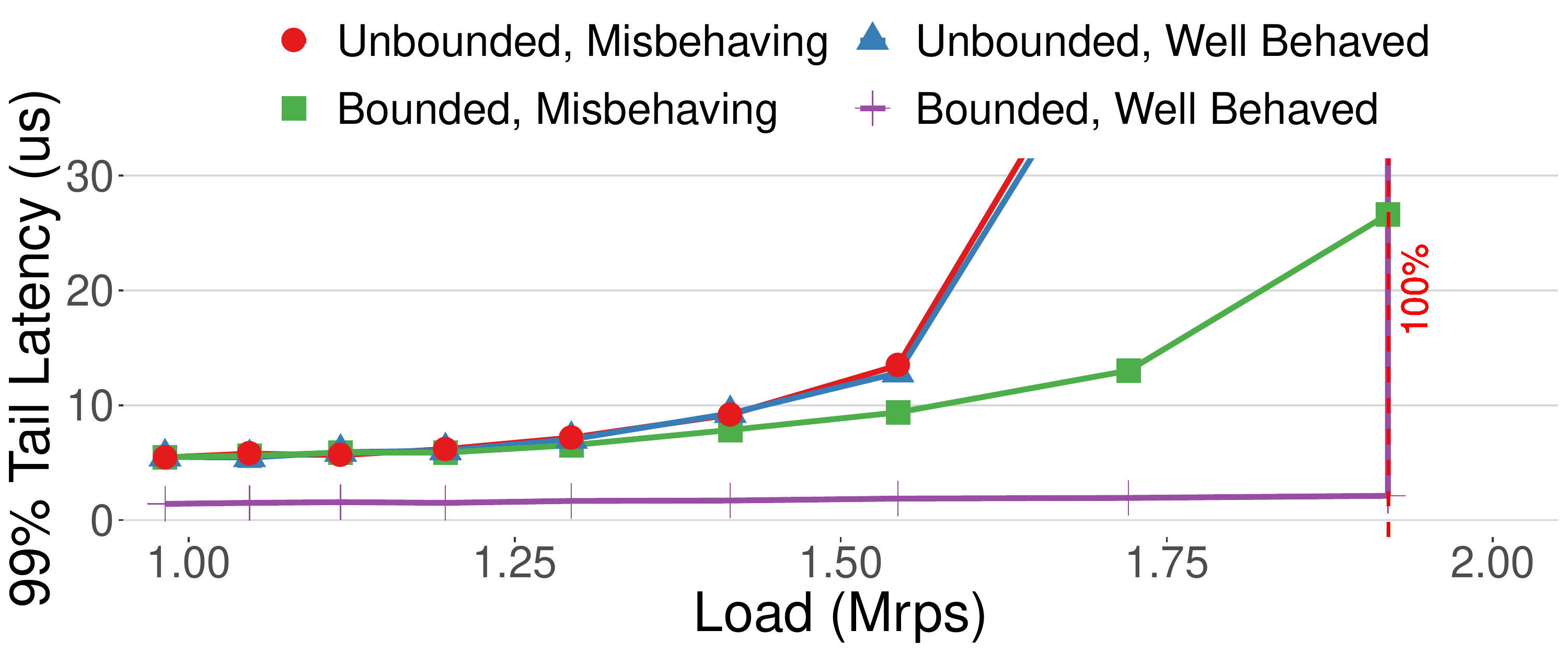}
  \caption{99\% tail latency vs load for both the well-behaved and misbehaved threads during two experiments: one in which bounded message processing time is enabled and one in which it is disabled.}
  \label{fig:bounded-scheduling-eval}
\end{figure}

Figure~\ref{fig:bounded-scheduling-eval} shows the 99\% tail latency vs load for the well-behaved and misbehaving threads for the following two experiments:
\begin{itemize}
    \item {\bf Bounded time:} the bounded message processing time feature of the \name{} thread scheduler is enabled. 
    If a priority 0 thread takes longer than 1$\mu$s to process a request then its priority is lowered to priority 1.
    \item {\bf Unbounded time:} the bounded message processing time feature of the \name{} thread scheduler is disabled so both threads remain at the same priority level and all requests are processed by the core in FIFO order.
\end{itemize}

We expect an application with at most one message at a time in the RX queue, to have a tail latency bounded by $2 \cdot 43ns + 13ns + 2 \cdot 1000ns + 50ns = 2.15 \mu s$. 
This matches our experiments: With the bounded message processing time feature enabled, the tail latency of the well-behaved thread never exceeds 2.1$\mu$s, until the offered load on the system exceeds 100\% (1.9 Mrps). 
This is despite using a Poisson arrival process that will occasionally allow more than one message in the RX queue.
The \name{} lowers the priority of the misbehaving application the first time that it takes longer than 1$\mu$s to process a request.
Hence, the well-behaved application quickly becomes strictly higher priority than the misbehaving application and its requests are never trapped behind a long 5$\mu$s request.
Note that the bounded message processing time approach can sustain higher loads because, by processing shorter requests first, it keeps the queues smaller.
The feature is therefore shown to strictly bound the tail latency for high priority applications.

\paragraph{e. Core selection algorithm:}
\label{sec:eval-load-balance}
The hardware core selection algorithm steers incoming messages to \name{} cores for processing. 
We evaluate and compare three different algorithms using a workload representative of an application like Redis~\cite{redis}. 
We assume that 99.5\% of messages are simple get/put requests (modeled by a \name{} service time of 500ns) and 0.5\% of messages are complex range queries (with a 5$\mu$s service time).
We compare three core selection techniques:
\begin{itemize}
    \item {\bf RSS (Receive Side Scaling):} This is a simple load-balancing algorithm commonly used by modern NICs. 
    One thread runs on each core and is fed by a separate global RX queue (one per-thread, which is also one per-core). 
    Each thread is assigned a unique port number, and the load generator selects a port number uniformly at random.
    \item {\bf JBSQ:} This is the algorithm described in Section~\ref{sec:design-load-balancing}. 
    We run one thread per core, allocate one global RX queue for all threads (\ie, all threads share the same port number).  
    The JBSQ algorithm load-balances requests to cores.
    \item {\bf JBSQ-PRE:} In this prioritized version, the short requests are assigned priority 0 (high), and long requests run at priority 1 (low). 
    Each type of request has its own port number. 
    We run two threads on each core (one per-priority) and run JBSQ with strict priority thread scheduling at each core as new messages arrive (described in Section~\ref{sec:design-thread-scheduling}).
\end{itemize}

\begin{figure}
  \includegraphics[width=\linewidth]{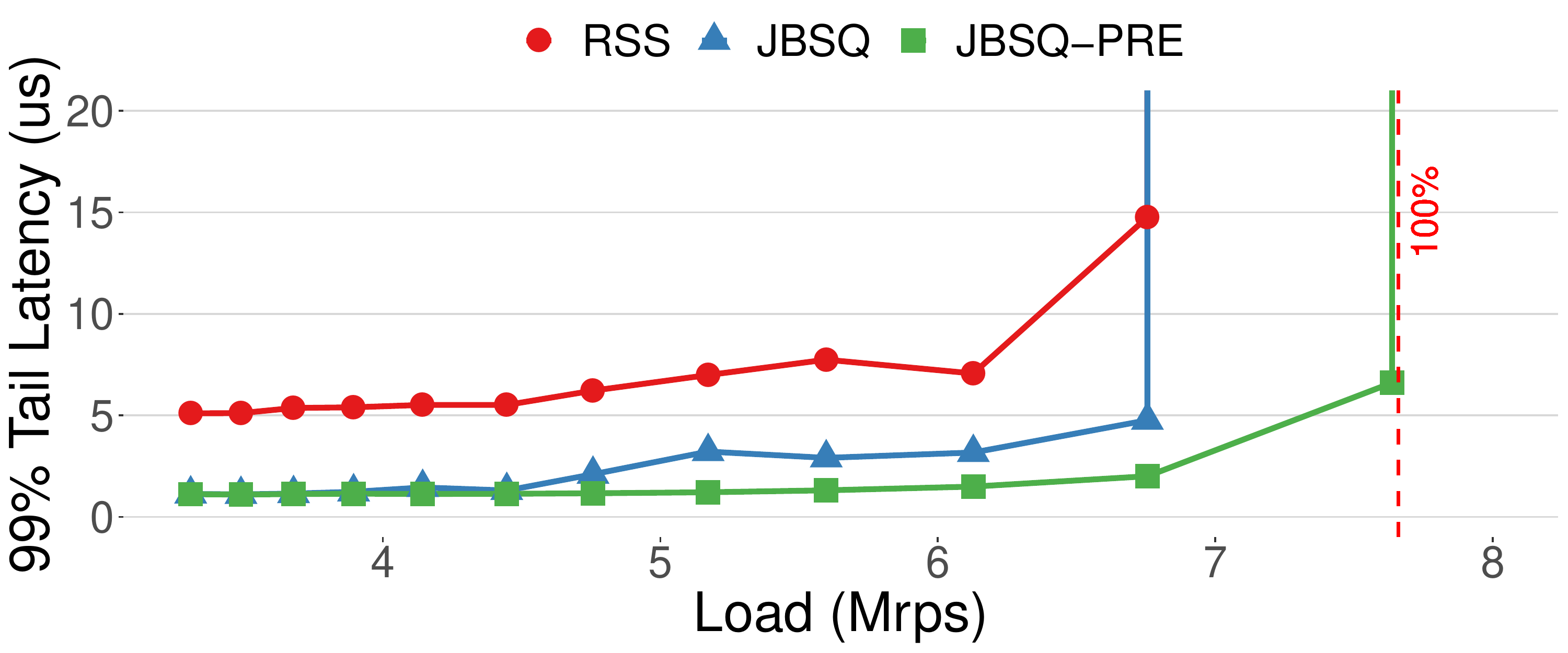}
  \caption{99\% tail latency vs load for three core-selection algorithms: RSS, JBSQ and JSBQ-PRE (with two priorities). Bimodal message service times: 99.5\% - 500ns, 0.5\% - 5$\mu$s.}
  \label{fig:load-balance-eval}
\end{figure}

Figure~\ref{fig:load-balance-eval} shows the 99\% tail latency vs load for the three techniques described above.
The tail latency of JBSQ is less than RSS because short requests do not get stuck behind long requests, unless all cores are busy processing long requests.
In that case, JBSQ-PRE is even better, because the \name{} thread scheduler will strictly prioritize processing short requests over long requests, preemptively if necessary.
JBSQ-PRE sustains higher overall load (almost 100\%) because it keeps the queues smaller by processing short requests first.

Our evaluation shows that with the combination of an efficient core selection algorithm and a fast per-message, preemptive, prioritized thread scheduling algorithm, we can sustain very high load and low latency from the \name{} cores. 

\paragraph{f. NDP transport:}
If the \name{} and extremely fine-grained computing become prevalent, we can expect large amounts of incast, hence our choice of NDP. 
We therefore evaluate our NDP implementation by running an 80-to-1 incast experiment. 
The experiment runs on 81 AWS FPGAs simulating 81 \name{}s with a total of 324 cores; the experiment is coordinated by Firesim.
The 81 \name{}s connect to a single switch via 200~Gb/s links; the RTT of the network is 3$\mu$s.
All 80 clients send a single 1024B message (in a 1088B packet) to the server at the same time. 
The bottleneck queue size is 81KB, and is therefore only large enough to hold 74 of the 80 packets; therefore, most of the packets will be queued, while others will be trimmed (when we enable NDP) or dropped (otherwise).
We run two experiments, one with NDP congestion control enabled and one with it disabled (by disabling packet trimming in the switch).

\begin{figure}
  \includegraphics[width=\linewidth]{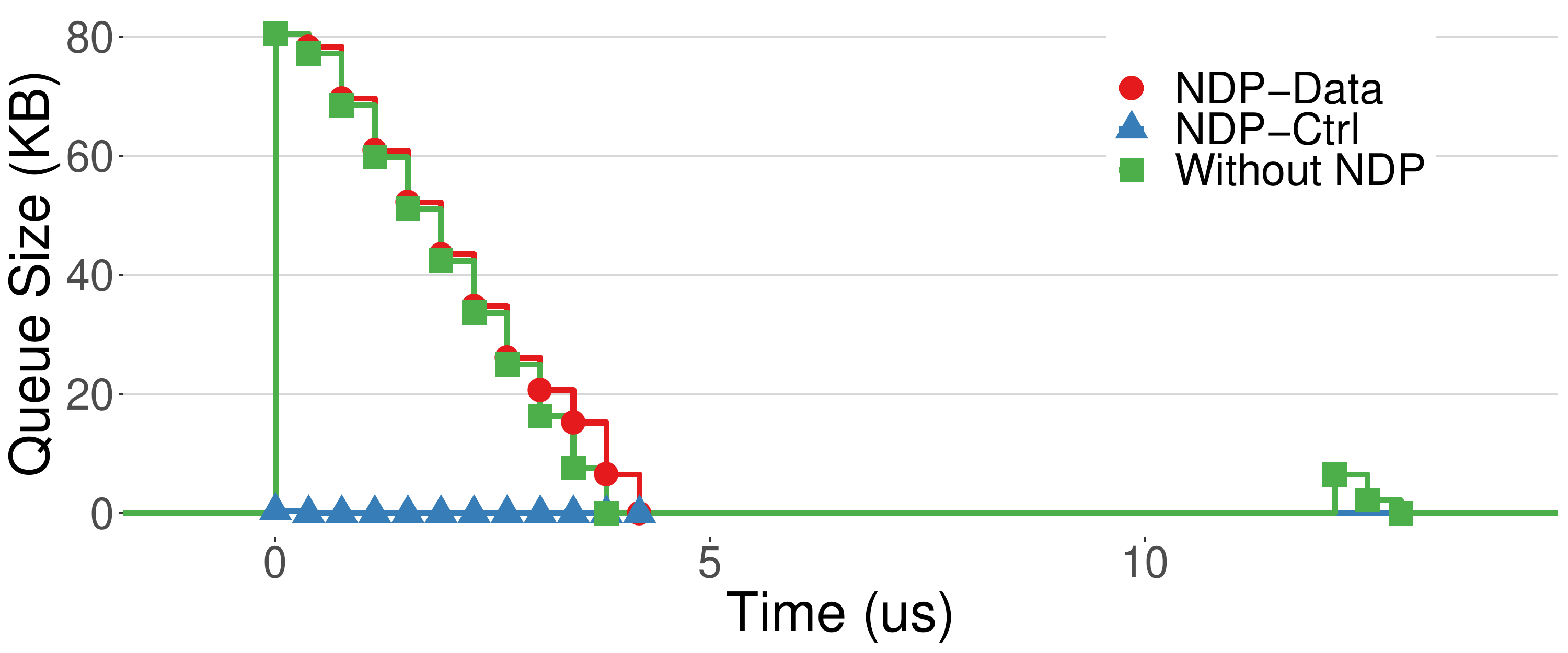}
  \caption{Occupancy of the bottleneck queue in the switch for 80-to-1 incast experiment, with and without NDP enabled.}
  \label{fig:incast-qsize}
\end{figure}

Figure~\ref{fig:incast-qsize} shows a time series of the occupancy of the bottleneck queue at the switch, with and without NDP enabled. 
At the beginning, we see all 80 packets arrive at the same time and filling up the switch queue.
Without NDP (green line), six packets are silently dropped at the onset of the incast.
The senders must infer that their packets were dropped using a timeout.
All of the retransmitted packets arrive at the same time, causing a smaller secondary incast. 
After 13$\mu$s the final byte of the final packet arrives.

On the other hand, with NDP enabled, six packets are trimmed and their headers are placed into the control queue and forwarded with high priority.
For each {\tt TRIM} packet received, the server generates a {\tt NACK} packet and a paced {\tt PULL} packet to tell the client to retransmit the dropped packet. 
{\tt PULL} packets are scheduled so that the retransmitted packets arrive at the bottleneck link at line-rate.
In total, it takes 4.2$\mu$s for the final byte of the final packet to be serialized onto the bottleneck link, which is about three times quicker than without NDP enabled.

\subsection{Application Benchmarks}
\paragraph{a. Key-value store:}
We ported the MICA key-value store~\cite{mica} to run on our quad-core \name{} prototype. 
This required minimal changes to the MICA source code; we modified 36 lines of functional code.

To compare with Nebula, our evaluation stores 10k key-value pairs (16B keys and 512B values) on each core. 
Each core holds distinct key ranges, obviating the need for inter-core synchronization.
The load generator sends a 50:50 mix of read/write queries with keys picked uniformly from the set. 

\begin{figure}
  \includegraphics[width=\linewidth]{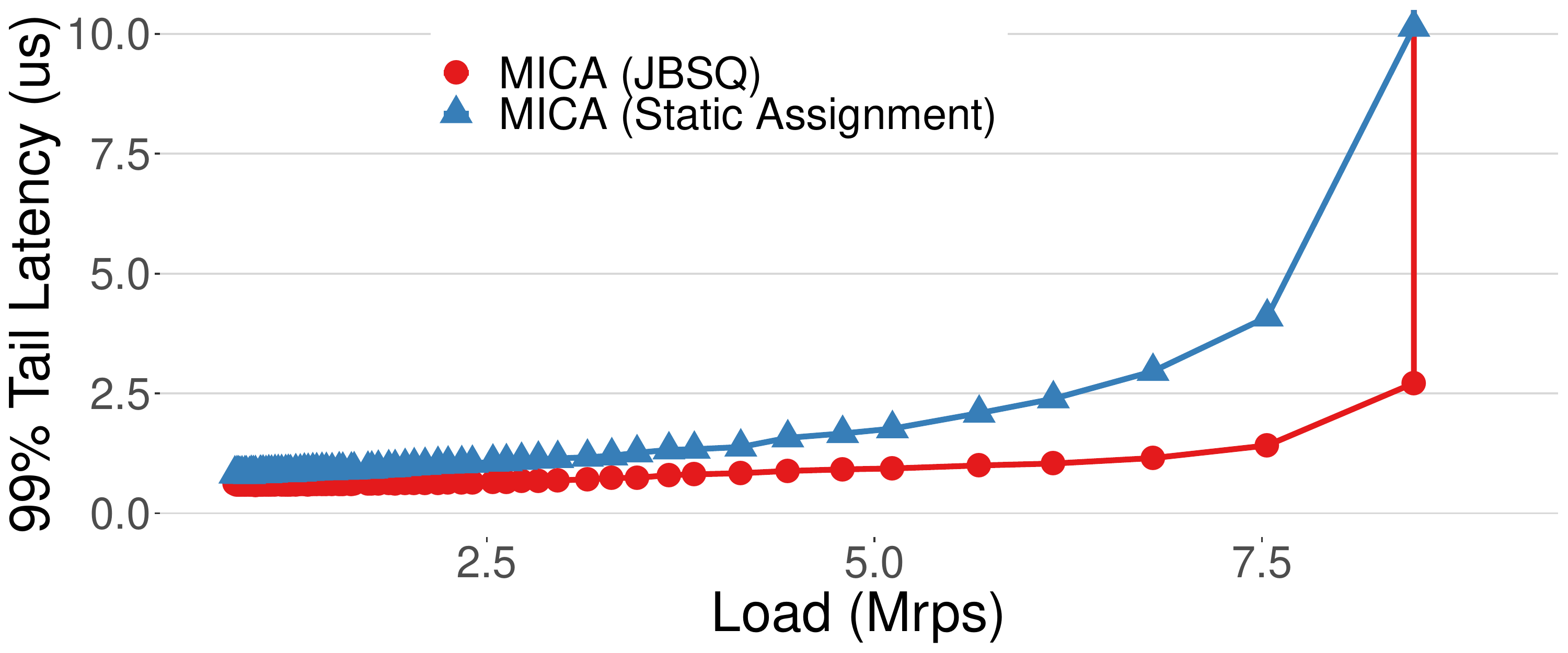}
  \caption{MICA key-value store: 99\% wire-to-wire tail latency vs load for READ and WRITE requests. 
  The latency includes an additional 43ns wire latency to/from the load-generator.}
  \label{fig:mica-eval}
\end{figure}

Figure~\ref{fig:mica-eval} compares the 99\% tail latency vs load for two different core-selection policies: JBSQ and static core assignment.
While using JBSQ core selection for this workload leads to incorrect application behavior (since each request must be serviced by a fixed core dictated by the key), we include it to match how Nebula was evaluated.

It is an ambitious exercise for our tiny Rocket cores to compete with Nebula's much beefier out-of-order, triple-issue, ARM cores running at 2~GHz, with a 16MB LLC, and 45ns DRAM access time. 
Nonetheless, we see that the \name{} performs better. 
At low load, \name{} running JBSQ leads to a 592ns 99\% tail latency for MICA, and 823ns for static core assignment --- including the link latency between the load generator and the \name.
Nebula reports a ``sub-$2\mu$s 99\% tail latency.''
Nebula is able to achieve a maximum per-core load of about 1.5~Mrps, whereas the \name{} is higher at 2.1~Mrps.

The main takeaway is that the \name{}, like Nebula, eliminates memory-bandwidth interference and therefore achieves similar low tail-latency and high throughput, but with a less powerful core.
The systems use different methods to eliminate memory-bandwidth interference.
Nebula limits the amount of LLC space allocated to a particular application based on its average service time, which breaks down when the request processing time is unknown or highly variable. 
On the other hand, the \name{} uses a separate hardware FIFOs for network data and is unaffected by variations in request processing time.

% Using a single receive queue to load balance requests across cores, Nebula reports a mean zero-load latency of \textasciitilde{}630ns for receiving and processing the request. We measured a mean latency of \textasciitilde{}851ns wire-to-wire, which also includes sending the response. The red line in Figure~\ref{fig:mica-eval} shows the 99\% tail latency vs. load for one queue for all cores (same setup as Nebula). For correct application behavior, however, cores should have separate RX queues, as cores are responsible for a different subsets of keys. With separate RX queues, the tail latency is higher (blue line), because the requests are not load balanced across the cores.

\paragraph{b. Chain replication:}
Strongly consistent, fault-tolerant key-value stores often use chain replication. 
We evaluate the chain replication protocol from NetChain~\cite{netchain} by porting it to run on multiple \name{}s, running on top of our MICA key-value store.  
To WRITE to the store, the client sends a request to the first replica in the chain. 
The replica applies the WRITE and forwards the request to the next replica in the chain, as indicated in the packet. 
The last replica sends an ACK to the client to complete the WRITE. 
READ requests are sent directly to the last node in the chain, which replies directly to the client.

To compare to the NetChain evaluation, we used 10k 16B/64B key/value pairs, and measure the end-to-end latency for a 3-way replicated WRITE from the client through the servers and back. We configure the switch that is connecting all four hosts to forward packets with zero latency and use 43ns link latencies.
With three replicas, NetChain reports 9.7$\mu$s mean zero-load latency. 
We measured a mean latency of 1.1$\mu$s and a 99\% tail latency of just 1.5$\mu$s on the \name{}.  
The \name{} client took only 130ns, compared to approximately 3-4$\mu$s for the DPDK client used in the NetChain evaluation. 
NetChain is implemented using programmable switches, hence the deployment is limited in terms of key-value size and capacity constraints, lack of congestion-control/reliable transport, and it relies on an external failure detector.
Our implementation does not suffer from these same limitations because the \name{} is a general-purpose processor, which is more flexible than a programmable switch.

\section{Discussion}
%\epigraph{\vspace{-1em}``Perfection is achieved, not when there is nothing more to add, but when there is nothing left to take away.''}{Antoine de Saint-Exup\'{e}ry, 1942\vspace{-1.2em}}

The \name{} is deliberately simple: We believe that minimizing latency requires us to strip away complexity, and move what we can to pipelined hardware. 
Our prototype is also simple because of the constraints of a university research project. 
The designer of a complete \name{} will need to consider additional key tradeoffs.

\paragraph{GPRs vs CSRs.} The \name{} prototype repurposes two GPRs for the head ({\tt netRX}) and tail ({\tt netTX}) of the network queues. 
In a CPU with 32 registers, we can likely afford to lose two; our evaluated applications did not appear to suffer. 
However, we also experimented with a design where the head and tail registers are implemented using control status registers (CSRs) instead, an alternative to consider if GPRs are limited. 

\paragraph{In-order execution.} The \name{} prototype is based on a simple 5-stage, in-order RISC-V Rocket core.  
While our prototype required very minor modifications to the CPU pipeline, if an out-of-order processor is used, more invasive changes would be required to ensure that words read from the RX queue are in FIFO order.

\paragraph{Floating-point operations.} Our prototype places messages directly into the \emph{integer} register file.
Scientific computing applications use \emph{floating point} arithmetic, and floating point GPRs. The RISC-V ISA~\cite{riscv-isa} includes instructions to copy bits directly between integer and floating point registers, although we have not used them.

% \paragraph{Kernel} Our prototype replaces Linux with our own minimal operating system, the nanokernel, designed to minimize average and tail latency for context switches. Our nanokernel currently lacks virtual memory and different privilege modes;
% applications and the nanokernel currently execute in the same address space and at the same privilege level. Our next generation nanokernel will address these issues, and we do not anticipate much increase in latency.

\section{Related Work}
% ``The hunt for the killer microseconds''~\cite{killer-microseconds} is forcing the research community to rethink how we build distributed systems.

\paragraph{Low-latency software systems for RPC.}
Recent work on low-latency RPC systems focuses on load balancing requests across cores by approximating a single-queue system using work-stealing (like ZygOS~\cite{zygos}) or preempting requests at microsecond timescales (Shinjuku~\cite{shinjuku}) to avoid head-of-line blocking and to manage requests with highly-variable service times.
However, the inter-core synchronization and software preemption incurs non-negligible overheads and degrade performance.
Software thread scheduling and core selection are too slow and coarse-grained for sub-microsecond RPCs. 
Therefore, the \name{} implements them both in hardware leading to significantly higher performance.

Unlike the \name{}, eRPC~\cite{eRPC} takes the other extreme and runs everything in software, and through clever optimizations, achieves impressively low latency on a commodity server for the common case. 
eRPC has good average performance, but its common-case optimizations sacrifice tail latency, which often dictate application performance. 
The \name{}'s hardware pipeline and direct coupling to the register file makes median and tail RPC latencies almost identical. 

\paragraph{Hardware extensions for RPC systems.}
Others have proposed implementing core selection algorithms in hardware.
RPCvalet~\cite{rpcvalet} and Nebula~\cite{nebula} are both built on top of the Scale-out NUMA architecture~\cite{scale-out-numa}.
RPCvalet implements a single queue system, which in theory provides optimal performance. 
However, it ran into memory bandwidth contention issues, which they later resolve in Nebula.
Both Nebula and R2P2~\cite{r2p2} implement the JBSQ load balancing policy; Nebula runs JBSQ on the server whereas R2P2 runs JBSQ in a programmable switch.
Like Nebula, the \name{} also implements JBSQ to steer requests to cores.

%\paragraph{RDMA-based RPC systems.}
RDMA is designed to give direct access to a remote server's memory.
Many NICs now offer RDMA in hardware and can respond in a few microseconds.
Several systems such as HERD~\cite{herd}, FaSST~\cite{fasst}, and DrTM+R~\cite{drtmr} exploit RDMA to build applications and services on top.
However, the \name{} targets RPC-based applications that need low latency access to remote CPUs, {\em not} remote memory.

%\paragraph{SmartNICs.}
SmartNICs (NICs with CPUs on them) are in deployment today by cloud service providers (CSPs)~\cite{nitro, bluefield, pensando}, to offload infrastructure software from the main server to CPUs on the NIC. 
However, these may actually increase the RPC latency, unless they adopt \name{}-like designs on the NIC.  

\paragraph{Transport protocols in hardware.}
We are not the first to implement the transport layer and congestion control in hardware. 
Modern NICs that support RDMA over Converged Ethernet (RoCE) already implement DCQN~\cite{dcqcn} in hardware. 
In the academic research community, Tonic~\cite{tonic} proposes a framework for implementing congestion control in hardware.
The \name{}'s NDP implementation draws upon ideas in Tonic, and goes further to build and evaluate a full system.

\paragraph{Register file interface.}
GPRs were first used by the J-machine~\cite{jmachine} for low-latency inter-core communication on the same machine, but were abandoned because of the difficulty in isolating threads running on the same core. 
Moore's Law has helped: From $10^6$ transistors per chip in 1989 to over $10^{10}$ today, \name{} can afford per-thread local queues, not feasible at the time of the J-machine. 

% Moreover, the \name{} targets network data as opposed to application data that exhibits high spatial-and-temporal locality, hence, limiting the impact of having a direct core-to-core register interface---most requests are handled using the L1, L2, and LLC caches.
% Network data doesn't exhibit such locality and traffic always flows through the register interface.
\section{Conclusion}
Today's RISC CPUs are optimized for load-store operations to and from memory. Memory data is treated as a first-class citizen. 
But modern workloads frequently process huge numbers of packets, \eg, RPCs for distributed applications and stream processing for NFV. 
Rather than burden packets with traversing a hierarchy optimized for data sitting in memory, we propose providing them with a new optimized fast path, directly into the heart of the CPU. 
Hence, we aim to elevate packet data to the same importance as memory data.

We set out to accelerate distributed applications by minimizing RPC tail latency. As  applications employ more parallelism, the RPC fanout increases, and so response time is increasingly determined by tail, not median, latency. The bottom line is that, by placing essential functions into hardware, we have driven the RPC tail-latency much closer to the median, potentially accelerating large distributed applications by an order of magnitude. 
\bibliographystyle{plain}
\bibliography{references}

\begin{thebibliography}{10}

\bibitem{f1}
Amazon ec2 f1 instances.
\newblock \url{https://aws.amazon.com/ec2/instance-types/f1/}.
\newblock Accessed on 2020-08-10.

\bibitem{sprocket}
Lixiang Ao, Liz Izhikevich, Geoffrey~M. Voelker, and George Porter.
\newblock {Sprocket: A Serverless Video Processing Framework}.
\newblock In {\em ACM SoCC}, 2018.

\bibitem{tonic}
Mina~Tahmasbi Arashloo, Alexey Lavrov, Manya Ghobadi, Jennifer Rexford, David
  Walker, and David Wentzlaff.
\newblock Enabling programmable transport protocols in high-speed nics.
\newblock In {\em 17th $\{$USENIX$\}$ Symposium on Networked Systems Design and
  Implementation ($\{$NSDI$\}$ 20)}, pages 93--109, 2020.

\bibitem{microservices}
Michael Armbrust, Armando Fox, Rean Griffith, Anthony~D. Joseph, Randy~H. Katz,
  Andrew Konwinski, Gunho Lee, David~A. Patterson, Ariel Rabkin, Ion Stoica,
  and Matei Zaharia.
\newblock Above the clouds: A berkeley view of cloud computing.
\newblock Technical Report UCB/EECS-2009-28, EECS Department, University of
  California, Berkeley, Feb 2009.

\bibitem{rocket-chip}
Krste Asanovic, Rimas Avizienis, Jonathan Bachrach, Scott Beamer, David
  Biancolin, Christopher Celio, Henry Cook, Daniel Dabbelt, John Hauser, Adam
  Izraelevitz, et~al.
\newblock The rocket chip generator.
\newblock {\em EECS Department, University of California, Berkeley, Tech. Rep.
  UCB/EECS-2016-17}, 2016.

\bibitem{nitro}
Aws nitro system.
\newblock \url{https://aws.amazon.com/ec2/nitro/}.
\newblock Accessed on 02/04/2020.

\bibitem{chisel}
Jonathan Bachrach, Huy Vo, Brian Richards, Yunsup Lee, Andrew Waterman, Rimas
  Avi{\v{z}}ienis, John Wawrzynek, and Krste Asanovi{\'c}.
\newblock Chisel: constructing hardware in a scala embedded language.
\newblock In {\em DAC Design Automation Conference 2012}, pages 1212--1221.
  IEEE, 2012.

\bibitem{rdma-networks}
Carsten Binnig, Andrew Crotty, Alex Galakatos, Tim Kraska, and Erfan Zamanian.
\newblock The end of slow networks: It’s time for a redesign.
\newblock {\em Proc. VLDB Endow.}, 9(7):528–539, March 2016.

\bibitem{RMT}
Pat Bosshart, Glen Gibb, Hun-Seok Kim, George Varghese, Nick McKeown, Martin
  Izzard, Fernando Mujica, and Mark Horowitz.
\newblock Forwarding metamorphosis: Fast programmable match-action processing
  in hardware for sdn.
\newblock {\em ACM SIGCOMM Computer Communication Review}, 43(4):99--110, 2013.

\bibitem{cirrus}
Joao Carreira, Pedro Fonseca, Alexey Tumanov, Andrew Zhang, and Randy Katz.
\newblock {Cirrus: A Serverless Framework for End-to-End ML Workflows}.
\newblock In {\em ACM SoCC}, 2019.

\bibitem{drtmr}
Yanzhe Chen, Xingda Wei, Jiaxin Shi, Rong Chen, and Haibo Chen.
\newblock Fast and general distributed transactions using rdma and htm.
\newblock In {\em Proceedings of the Eleventh European Conference on Computer
  Systems}, pages 1--17, 2016.

\bibitem{rpcvalet}
Alexandros Daglis, Mark Sutherland, and Babak Falsafi.
\newblock Rpcvalet: Ni-driven tail-aware balancing of $\mu$s-scale rpcs.
\newblock In {\em Proceedings of the Twenty-Fourth International Conference on
  Architectural Support for Programming Languages and Operating Systems}, pages
  35--48, 2019.

\bibitem{jmachine}
William~J Dally, Andrew Chien, Stuart Fiske, Waldemar Horwat, and John Keen.
\newblock The j-machine: A fine grain concurrent computer.
\newblock Technical report, MASSACHUSETTS INST OF TECH CAMBRIDGE MICROSYSTEMS
  RESEARCH CENTER, 1989.

\bibitem{context-switch-overhead}
Francis~M. David, Jeffrey~C. Carlyle, and Roy~H. Campbell.
\newblock Context switch overheads for linux on arm platforms.
\newblock In {\em Proceedings of the 2007 Workshop on Experimental Computer
  Science}, ExpCS '07, page 3–es, New York, NY, USA, 2007. Association for
  Computing Machinery.

\bibitem{davidson2018celerity}
Scott Davidson, Shaolin Xie, Christopher Torng, Khalid Al-Hawai, Austin
  Rovinski, Tutu Ajayi, Luis Vega, Chun Zhao, Ritchie Zhao, Steve Dai, et~al.
\newblock The celerity open-source 511-core risc-v tiered accelerator fabric:
  Fast architectures and design methodologies for fast chips.
\newblock {\em IEEE Micro}, 38(2):30--41, 2018.

\bibitem{ddio}
Intel corporation. intel data direct i/o technology (intel ddio): A primer.
\newblock
  \url{https://www.intel.com/content/dam/www/public/us/en/documents/technology-briefs/data-direct-i-o-technology-brief.pdf}.
\newblock Accessed on 2020-08-17.

\bibitem{tail-at-scale}
Jeffrey Dean and Luiz~Andr{\'e} Barroso.
\newblock The tail at scale.
\newblock {\em Communications of the ACM}, 56(2):74--80, 2013.

\bibitem{nanopu-github}
{nanoPU GitHub}.
\newblock \url{https://github.com/l-nic}.
\newblock Accessed on 08/17/2020.

\bibitem{ExCamera}
Sadjad Fouladi, Riad~S Wahby, Brennan Shacklett, Karthikeyan~Vasuki
  Balasubramaniam, William Zeng, Rahul Bhalerao, Anirudh Sivaraman, George
  Porter, and Keith Winstein.
\newblock {Encoding, Fast and Slow: Low-latency Video Processing Using
  Thousands of Tiny Threads}.
\newblock In {\em USENIX NSDI}, 2017.

\bibitem{ndp}
Mark Handley, Costin Raiciu, Alexandru Agache, Andrei Voinescu, Andrew~W Moore,
  Gianni Antichi, and Marcin W{\'o}jcik.
\newblock Re-architecting datacenter networks and stacks for low latency and
  high performance.
\newblock In {\em Proceedings of the Conference of the ACM Special Interest
  Group on Data Communication}, pages 29--42, 2017.

\bibitem{dca}
Ram Huggahalli, Ravi Iyer, and Scott Tetrick.
\newblock Direct cache access for high bandwidth network i/o.
\newblock In {\em Proceedings of the 32nd Annual International Symposium on
  Computer Architecture}, ISCA ’05, page 50–59, USA, 2005. IEEE Computer
  Society.

\bibitem{event-driven-pisa}
Stephen Ibanez, Gianni Antichi, Gordon Brebner, and Nick McKeown.
\newblock Event-driven packet processing.
\newblock In {\em Proceedings of the 18th ACM Workshop on Hot Topics in
  Networks}, pages 133--140, 2019.

\bibitem{lnic}
Stephen Ibanez, Muhammad Shahbaz, and Nick McKeown.
\newblock The case for a network fast path to the cpu.
\newblock In {\em Proceedings of the 18th ACM Workshop on Hot Topics in
  Networks}, pages 52--59, 2019.

\bibitem{caribou}
Zsolt Istv\'{a}n, David Sidler, and Gustavo Alonso.
\newblock Caribou: Intelligent distributed storage.
\newblock {\em Proc. VLDB Endow.}, 10(11):1202–1213, August 2017.

\bibitem{hardware-consensus}
Zsolt Istv\'{a}n, David Sidler, Gustavo Alonso, and Marko Vukolic.
\newblock Consensus in a box: Inexpensive coordination in hardware.
\newblock In {\em Proceedings of the 13th Usenix Conference on Networked
  Systems Design and Implementation}, NSDI’16, page 425–438, USA, 2016.
  USENIX Association.

\bibitem{netchain}
Xin Jin, Xiaozhou Li, Haoyu Zhang, Nate Foster, Jeongkeun Lee, Robert
  Soul{\'e}, Changhoon Kim, and Ion Stoica.
\newblock Netchain: Scale-free sub-rtt coordination.
\newblock In {\em 15th $\{$USENIX$\}$ Symposium on Networked Systems Design and
  Implementation ($\{$NSDI$\}$ 18)}, pages 35--49, 2018.

\bibitem{shinjuku}
Kostis Kaffes, Timothy Chong, Jack~Tigar Humphries, Adam Belay, David
  Mazi{\`e}res, and Christos Kozyrakis.
\newblock Shinjuku: Preemptive scheduling for $\mu$second-scale tail latency.
\newblock In {\em 16th $\{$USENIX$\}$ Symposium on Networked Systems Design and
  Implementation ($\{$NSDI$\}$ 19)}, pages 345--360, 2019.

\bibitem{eRPC}
Anuj Kalia, Michael Kaminsky, and David Andersen.
\newblock Datacenter rpcs can be general and fast.
\newblock In {\em 16th $\{$USENIX$\}$ Symposium on Networked Systems Design and
  Implementation ($\{$NSDI$\}$ 19)}, pages 1--16, 2019.

\bibitem{herd}
Anuj Kalia, Michael Kaminsky, and David~G Andersen.
\newblock Using rdma efficiently for key-value services.
\newblock In {\em Proceedings of the 2014 ACM conference on SIGCOMM}, pages
  295--306, 2014.

\bibitem{fasst}
Anuj Kalia, Michael Kaminsky, and David~G Andersen.
\newblock Fasst: Fast, scalable and simple distributed transactions with
  two-sided ($\{$RDMA$\}$) datagram rpcs.
\newblock In {\em 12th $\{$USENIX$\}$ Symposium on Operating Systems Design and
  Implementation ($\{$OSDI$\}$ 16)}, pages 185--201, 2016.

\bibitem{firesim}
Sagar Karandikar, Howard Mao, Donggyu Kim, David Biancolin, Alon Amid, Dayeol
  Lee, Nathan Pemberton, Emmanuel Amaro, Colin Schmidt, Aditya Chopra, et~al.
\newblock Firesim: Fpga-accelerated cycle-exact scale-out system simulation in
  the public cloud.
\newblock In {\em 2018 ACM/IEEE 45th Annual International Symposium on Computer
  Architecture (ISCA)}, pages 29--42. IEEE, 2018.

\bibitem{INT}
Changhoon Kim, Anirudh Sivaraman, Naga Katta, Antonin Bas, Advait Dixit, and
  Lawrence~J Wobker.
\newblock In-band network telemetry via programmable dataplanes.
\newblock In {\em ACM SIGCOMM}, 2015.

\bibitem{r2p2}
Marios Kogias, George Prekas, Adrien Ghosn, Jonas Fietz, and Edouard Bugnion.
\newblock R2p2: Making rpcs first-class datacenter citizens.
\newblock In {\em 2019 $\{$USENIX$\}$ Annual Technical Conference
  ($\{$USENIX$\}$$\{$ATC$\}$ 19)}, pages 863--880, 2019.

\bibitem{mica}
Hyeontaek Lim, Dongsu Han, David~G Andersen, and Michael Kaminsky.
\newblock $\{$MICA$\}$: A holistic approach to fast in-memory key-value
  storage.
\newblock In {\em 11th $\{$USENIX$\}$ Symposium on Networked Systems Design and
  Implementation ($\{$NSDI$\}$ 14)}, pages 429--444, 2014.

\bibitem{rdma-shuffle}
Feilong Liu, Lingyan Yin, and Spyros Blanas.
\newblock Design and evaluation of an rdma-aware data shuffling operator for
  parallel database systems.
\newblock In {\em Proceedings of the Twelfth European Conference on Computer
  Systems}, EuroSys ’17, page 48–63, New York, NY, USA, 2017. Association
  for Computing Machinery.

\bibitem{bluefield}
Mellanox bluefield-2.
\newblock \url{https://www.mellanox.com/products/bluefield2-overview}.
\newblock Accessed on 02/04/2020.

\bibitem{homa}
Behnam Montazeri, Yilong Li, Mohammad Alizadeh, and John Ousterhout.
\newblock Homa: A receiver-driven low-latency transport protocol using network
  priorities.
\newblock In {\em Proceedings of the 2018 Conference of the ACM Special
  Interest Group on Data Communication}, pages 221--235, 2018.

\bibitem{pensando}
Naples dsc-100 distributed services card.
\newblock
  \url{https://www.pensando.io/assets/documents/Naples_100_ProductBrief-10-2019.pdf}.
\newblock Accessed on 02/04/2020.

\bibitem{scale-out-numa}
Stanko Novakovic, Alexandros Daglis, Edouard Bugnion, Babak Falsafi, and Boris
  Grot.
\newblock Scale-out numa.
\newblock {\em ACM SIGPLAN Notices}, 49(4):3--18, 2014.

\bibitem{gcc-options}
{Options for Code Generation Conventions}.
\newblock
  \url{https://gcc.gnu.org/onlinedocs/gcc/Code-Gen-Options.html#Code-Gen-Options}.
\newblock Accessed on 02/04/2020.

\bibitem{shenango}
Amy Ousterhout, Joshua Fried, Jonathan Behrens, Adam Belay, and Hari
  Balakrishnan.
\newblock Shenango: Achieving high $\{$CPU$\}$ efficiency for latency-sensitive
  datacenter workloads.
\newblock In {\em 16th $\{$USENIX$\}$ Symposium on Networked Systems Design and
  Implementation ($\{$NSDI$\}$ 19)}, pages 361--378, 2019.

\bibitem{ramcloud}
John Ousterhout, Arjun Gopalan, Ashish Gupta, Ankita Kejriwal, Collin Lee,
  Behnam Montazeri, Diego Ongaro, Seo~Jin Park, Henry Qin, Mendel Rosenblum,
  Stephen Rumble, Ryan Stutsman, and Stephen Yang.
\newblock The ramcloud storage system.
\newblock {\em ACM Trans. Comput. Syst.}, 33(3), August 2015.

\bibitem{zygos}
George Prekas, Marios Kogias, and Edouard Bugnion.
\newblock Zygos: Achieving low tail latency for microsecond-scale networked
  tasks.
\newblock In {\em Proceedings of the 26th Symposium on Operating Systems
  Principles}, pages 325--341, 2017.

\bibitem{redis}
Redis.
\newblock \url{https://redis.io/}.
\newblock Accessed on 2020-08-12.

\bibitem{riscv-isa}
Risc-v specifications.
\newblock \url{https://riscv.org/technical/specifications/}.
\newblock Accessed on 2020-08-17.

\bibitem{rocket-chip-github}
Rocket-chip github.
\newblock \url{https://github.com/chipsalliance/rocket-chip}.
\newblock Accessed on 08/17/2020.

\bibitem{nebula}
Mark Sutherland, Siddharth Gupta, Babak Falsafi, Virendra Marathe, Dionisios
  Pnevmatikatos, and Alexandros Daglis.
\newblock The nebula rpc-optimized architecture.
\newblock Technical report, 2020.

\bibitem{resQ}
Amin Tootoonchian, Aurojit Panda, Chang Lan, Melvin Walls, Katerina Argyraki,
  Sylvia Ratnasamy, and Scott Shenker.
\newblock Resq: Enabling slos in network function virtualization.
\newblock In {\em 15th $\{$USENIX$\}$ Symposium on Networked Systems Design and
  Implementation ($\{$NSDI$\}$ 18)}, pages 283--297, 2018.

\bibitem{verilator}
Verilator.
\newblock \url{https://www.veripool.org/wiki/verilator}.
\newblock Accessed on 2020-01-29.

\bibitem{netlock}
Zhuolong Yu, Yiwen Zhang, Vladimir Braverman, Mosharaf Chowdhury, and Xin Jin.
\newblock Netlock: Fast, centralized lock management using programmable
  switches.
\newblock In {\em Proceedings of the Annual Conference of the ACM Special
  Interest Group on Data Communication on the Applications, Technologies,
  Architectures, and Protocols for Computer Communication}, SIGCOMM ’20, page
  126–138, New York, NY, USA, 2020. Association for Computing Machinery.

\bibitem{dcqcn}
Yibo Zhu, Haggai Eran, Daniel Firestone, Chuanxiong Guo, Marina Lipshteyn,
  Yehonatan Liron, Jitendra Padhye, Shachar Raindel, Mohamad~Haj Yahia, and
  Ming Zhang.
\newblock Congestion control for large-scale rdma deployments.
\newblock {\em ACM SIGCOMM Computer Communication Review}, 45(4):523--536,
  2015.

\end{thebibliography}

\end{document}

% \title{The nanoPU: Minimizing RPCs for Massively-Parallel Applications}
\title{The \name{}: Making the Network Interface a First-Class Citizen \\ to Minimize RPC Tail Latency \\ (Appendices)}

\date{}
\maketitle

\thispagestyle{empty}

\appendix

\section{NDP Overview}
NDP~\cite{ndp} employs a few clever tricks to minimize end-to-end latency. 
First, packets are load-balanced over paths packet-by-packet (rather than flow-by-flow). 
This reduces congestion, but causes packet mis-sequencing; our prototype resequences packet data into correctly ordered messages. 
Second, NDP is receiver-driven; the receiving host decides when a sender can transmit packets, particularly during incast storms, by sending {\tt PULL} messages to allow the sender to transmit new {\tt DATA} packets or retransmit dropped packets. 
{\tt PULL} packets are paced to ensure that {\tt DATA} packets arrive at the bottleneck link at line-rate, avoiding further congestion. 
Third, if a packet encounters a full switch buffer, the header is trimmed and sent to the destination as a {\tt TRIM} packet; the packet data is dropped. 
The receiver responds to {\tt DATA} and {\tt TRIM} packets with {\tt ACKs} and {\tt NACKs}, respectively. 
Fourth, network switches forward control packets ({\tt PULL}, {\tt TRIM}, {\tt ACK}, {\tt NACK}) with high priority over {\tt DATA} packets.

\section{Design of the Message Buffer}
Here, we provide a brief overview of the buffer design we use in the \name{} prototype to perform message packetization and reassembly.

Our message buffer is divided into buffers of several different fixed sizes, and a free list for each size class keeps track of which buffers are available. When a buffer is allocated, the smallest available buffer that is large enough to store the whole message is selected.
For message reassembly, a buffer is allocated when the first packet of the message arrives from the network and is freed when the message is forwarded to the global RX queues.\footnote{An arriving packet is dropped at the ingress of the reassembly module if it is unable to allocate a buffer for the message.}
For message packetization, a buffer is allocated when the application writes the first word of the message from the core and is freed when the entire message has been ACKed by the receiver.

The design uses a table indexed by message identifier to keep track of where each message is stored (the buffer pointer), which simplifies out-of-order reassembly and retransmission.
Additionally, to find the position of a particular packet within the message, the hardware simply adds the appropriate offset to the message's buffer pointer.
For our evaluated workloads, we can configure the buffers such that the switch only drops packets when the total buffer space exceeds 96\% utilization.

\bibliographystyle{plain}
\bibliography{references}